\documentclass[traditabstract]{aa} 
\usepackage{graphicx,natbib}
\begin{document}
   \title{A diagnosis on torque reversals in 4U 1626$-$67}

   \subtitle{}

   \author{Z. Zhang\inst{1,2}
                 \and X.-D. Li\inst{1,2}
          }

   \institute{Department of Astronomy, Nanjing University, Nanjing 210093, China\\
         \and
             Key Laboratory of Modern Astronomy and Astrophysics (Nanjing University),
             Ministry of Education, China\\
             \email{zhangzhen19841209@gmail.com, lixd@nju.edu.cn}
             }

   \date{}

\def\lum{erg s$^{-1}$}

\newcommand{\ms}{M_\odot}

\abstract {Several X-ray pulsars have been observed to experience
torque reversals, which provide important observational clues to the
interaction between the neutron star magnetic field and the
accretion disk. We review the current models proposed for the torque
reversals and discuss their viability based on the observations of
the quasi-periodic oscillations (QPOs) in 4U 1626$-$67. Most of
these models seem to be incompatible with the evolution of the QPO
frequencies if they are interpreted in terms of the beat frequency
model. We suggest that winds or outflows from the neutron star and
the accretion disk may play an important role in accounting for the
spin-down in disk-fed neutron stars.

\keywords{accretion, accretion disk -- stars: neutron -- X-rays:
binaries -- X-rays: individual (4U 1626$-$67) -- X-rays: stars} }

\maketitle

\section{Introduction}
X-ray pulsars are magnetic neutron stars (NSs) in binary systems,
accreting from a normal companion star. They are excellent samples
to study the mass and angular momentum (AM) transfer between an NS
and the surrounding accretion flow \citep{nag89}. Mass accretion
onto an NS may occur through either an accretion disk fed by
Roche-lobe overflow, or capture of the stellar wind from the
companion. There has been long-term monitoring of the spin evolution
in X-ray pulsars, which is generally thought to result from the
interaction between the NS and the accretion flow. Especially, {\em
CGRO}/BATSE has provided continuous monitoring of X-ray pulsars
\citep{bil97}, among which 4U 1626$-$67 and GX 1$+$4 showed steady
spin-up until a sudden torque reversal appeared with the spin-down
rate similar to that in the spin-up stage but with opposite sign,
while Cen X$-$3, OAO 1657$-$415, and Her X$-$1 \citep{klo09}
exhibited a secular spin-up trend with short-time spin variations.

Obviously the torque reversals reflect a dramatic change in the
pattern of interaction between the NS magnetic fields and the accretion
disks. There have been lots of discussions in the literature on the
explanation of these intriguing  phenomena. However, a widely
accepted model has not emerged. In this work we focus on the torque
reversals in 4U 1626$-$67, a $7.66$ s X-ray pulsar. Its optical
counterpart was identified as KZ Tra, a faint blue star with little
or no reddening \citep{mcc77,bra83}. The short (42 minutes) orbital
period \citep{mid81} indicates that 4U 1626$-$67 is in an
ultra-compact binary with a hydrogen-depleted low-mass secondary
\citep{nel86}. Observations during 1977$-$1989 showed that
4U1626$-$67 was spinning up at a rate of $\dot{\nu}\simeq
8.54\times10^{-13}$ Hz\,s$^{-1}$. \citet{cha97a} found that, from
April 1991 to June 1996, the NS spun down at a rate of
$\dot{\nu}\simeq -7.17\times10^{-13}$ Hz\,s$^{-1}$. Though the torque
reversal was not observed directly, it was estimated to occur in
June 1990. More recent observations with {\em Fermi}/GBM and {\em
Swift}/BAT showed that 4U 1626$-$67 experienced a second torque
reversal near February  2008 \citep{cam10}. Since then it has been
following a steady spin-up at a mean rate of $\dot{\nu}\simeq
5\times10^{-13}$ Hzs$^{-1}$.

While the spin changes in X-ray pulsars provide one way to study NS
magnetic field-accretion disk interaction, quasi-periodic
oscillations (QPOs) offer another clue.  QPOs have been detected in
eight accreting X-ray pulsars \citep[][and references
therein]{boz09}, and are usually regarded as the signature of the
existence of an accretion disk. The measured QPO frequencies are in
the range of $\sim 1-200$ mHz, consistent with being relevant to the
inner radius of the accretion disk around a highly magnetized NS in
its bright X-ray state. In the most favored beat frequency model
\citep[BFM,][]{alp85,lam85}, the QPO frequency results from the beat
between the orbital frequency at the inner edge of the disk and the
spin frequency of the NS. Thus the QPO frequency may provide helpful
information about the accretion rate in the disk. In this paper,
after reviewing the current models of torque reversals in section 2,
we comment on the feasibility of these models based on the BFM
interpretation of the QPOs in 4U 1626$-$67 (section 3). In section 4
we present a new picture for the explanation of the torque reversals,
invoking winds or outflows from the NS and the accretion disk. We
summarize our results and discuss their possible implications in
section 5.

\section{A review of proposed interpretations for torque reversals}

\subsection{Magnetically threaded disk model}

The magnetically threaded disk model was first proposed and
investigated by \citet[][hereafter GL. See also \citet{boz09} for a
detailed description and discussion about the model]{gl79a,gl79b}.
Briefly speaking, in addition to the material torque
\begin{equation}
N_0=\dot{M}(GMr_0)^{1/2},
\end{equation}
where $M$ is the NS mass, $\dot{M}$ the accretion rate,  and $r_0$
the inner radius of the disk, GL introduced a magnetic torque,
\begin{equation}
N_{\rm mag}=-\int^{r_{\rm out}}_{r_0}{B_{\phi}B_{\rm z}r^2\,{\rm d}r},
\end{equation}
where $B_{\rm z}$ and $B_{\phi}$ are the vertical and toroidal
components of the magnetic field at the surfaces of the disk,
respectively,  $r_{\rm out}$ is the outer radius of the disk.
The magnetic torque is produced by the twist of the magnetic field lines
threading the disk. Thus, the pattern how the magnetic lines are
twisted decides the scale of the magnetic torque.
Assuming that the shear amplification of $B_{\phi}$ occurs on a
similar timescale to the reconnection of the slipped field lines, GL
and late investigations \citep[e.g.][]{wan95} showed that,
\begin{equation}
\frac{B_{\phi}}{B_{\rm z}}\propto \gamma(\Omega_{\rm s}-\Omega_{\rm K}),
\end{equation}
where $\gamma$ is the azimuthal pitch of the field lines,
$\Omega_{\rm s}$ and $\Omega_{\rm K}$ are the angular velocity of
the NS and the Keplerian angular velocity in the accretion disk,
respectively. Defining the corotation radius $r_{\rm c}$ as the
radius where the Keplerian angular velocity of the material in the
disk is the same as that of the NS, GL showed that the inner disk
with $r<r_{\rm c}$ spins the NS up, since $B_{\phi}B_{\rm z}<0$ and
$N_{\rm mag}>0$, while the part outside $r_{\rm c}$ draws the NS to
spin down. The total torque exerted on the NS is the sum of the
magnetic torque and the accretion torque, which can be described as,
\begin{equation}
N_{\rm GL}=\dot{M}(GMr_0)^{1/2}n(\omega_{\rm s}),
\end{equation}
where $n(\omega_{\rm s})$ is a dimensionless
function of the fastness parameter $\omega_s$,
\begin{equation}
\omega_{\rm s}\equiv\frac{\Omega_{\rm s}}{\Omega_K(r_0)}
\propto{\dot{M}^{-3/7}\Omega_{\rm s}B^{6/7}},
\end{equation}
where $B$ is the dipolar field strength on the surface of the NS.
The value of $n(\omega_s)$ can be either positive or negative, and there
is a critical value of $\omega_{\rm c}$, at which $n(\omega_{\rm
c})=0$, corresponding to an equilibrium state. If the NS is a slow
rotator, $\omega_{\rm s}<\omega_{\rm c}$, then $n(\omega_{\rm s})>0$, and
the NS experiences a spin-up torque; when $\omega_{\rm
c}<\omega_{\rm s}<1$, $n(\omega_{\rm s})$ becomes negative.

In the GL-type model, if $\dot{M}$ does not change much, given
sufficiently long time of evolution, the NS will always reach the
equilibrium state. The most straightforward explanation for the
torque reversals in X-ray pulsars is the change of $\dot{M}$.
However, this explanation has met several difficulties: (1) for Cen
X-3, a fine-tuned change of $\dot{M}$ is required \citep{bil97},
while observations suggested that the mass accretion rate did not
vary considerably during the torque reversals \citep{rai08}; (2) for
GX 1+4, the X-ray luminosity seemed to increase during the spin-down
episode \citep{cha97b}, which is in contrast with the prediction of
the original GL model.

In a modified version of the magnetically threaded disk model,
\citet{tor98} suggested that, in addition to the GL-type torque,
there is a torque coming from the interaction between the stellar
magnetic field and the disk's own, dynamo-generated magnetic field.
The azimuthal field strength $B_{\phi}$ is shown to be $\propto
r^{-3/4}$, decreasing with increasing $r$ more slowly than the
formalism $B_{\phi}\propto r^{-3}$ supposed by GL. Therefore, the
latter form of the magnetic torque dominate at large radii. The
total torque is then,
\begin{equation}
N_{\rm tot}=N_{\rm GL}\pm N_{\rm mag,dyn}.
\end{equation}
Whether the dynamo-generated magnetic torque is positive or negative
is not decidable, since the direction of the disk field is
arbitrary, and is not determined by the difference in angular
velocity between the star and the  disk as in GL, but a transition
between them is reflected as the torque reversal, if the magnetic
field in the disk reverses. Without invoking definite relation
between $\dot{M}$ and the spin state, Torkelsson (1998) indicated
that the timescale for an NS staying in one stable spin state is
uncertain, but the timescale for reversing the torque is the same as
that for reversing the magnetic field via diffusion, or the viscous
timescale. This might explain the short timescale for the reversals
compared with the long timescale over which the torque remains the
same as the difference between the diffusive (viscous) timescales
for the inner region of the disk and the entire accretion disk. The
complication of the model lies in numerous parameters, among which
quite a lot are unclear, so it is not certain how important the
dynamo-generated part is in the whole expression. Besides, this
model fails to explain the stability of the torque (over a timescale
of several years) in 4U 1626$-$67.

\subsection{Retrograde or warping disk model}

If a prograde disk provides the NS with positive torque, why cannot
a retrograde disk spins the NS down? In the simplest
situation, the torque the star receives comes from the accreted
material, so that
\begin{equation}
N=\pm{\dot{M}(GMr_0)^{1/2}},
\end{equation}
corresponding to a progarde and retrograde disk, respectively. A
transition between prograde and retrograde rotation of the disks
presents as the torque reversal naturally. Moreover, from Eq.~(7), a
similar spin-up and spin-down rate is anticipated, just coincident
with the observations of the spin evolutions of GX 1+4 and 4U 1626$-$67.

\citet{nel97} refreshed the idea of the retrograde disk for GX 1+4
\citep{mak88}, considering the fact that the X-ray luminosity
increases during the spin-down episode. As described above, with the
help of the retrograde disk, the X-ray luminosity can increase both
in the spin-up and spin-down episodes, which successfully avoids the
problem faced by the GL-type model.

While retrograde disks may exist in wind-fed systems like  GX 1$+$4
\citep{mat87,mat91,bor94,ruf99,kry05}, for Roche-lobe overflow
systems like 4U1626$-$67, the specific AM initially carried by the
accretion stream is comparable to the specific orbital AM of the
companion star, and should circularize in the prograde sense well
before reaching the NS magnetosphere (Lubow \& Shu 1975). This
problem might be solved by disk warping induced by X-ray irradiation
from the central source \citep{pri96} or by the magnetic torque
\citep{lai99}. If the inner part of the accretion disk flips over by
more than $90\degr$, and rotate in the opposite direction
\citep{vk98,wij99}, this would lead to a torque reversal. However,
it is highly uncertain whether irradiation can cause such a warping
and whether the warped disk can remain for a sufficiently long time.
More recent theoretical work and three-dimensional simulations
showed only slight ($\sim 10-20\%$) warping of the disk around
misaligned magnetic stars \citep{tp00,rom03}.

\subsection{Propeller model}

A general requirement for stable accretion from a disk to an NS is
that the velocity of the NS magnetosphere must be smaller than the
local Keplerian velocity at the inner edge of the disk, otherwise
the propeller mechanism will prohibit further accretion by ejecting
the accreted material away \citep{is75}. The propeller motion occurs
at the boundary of the magnetosphere, so the ejected material
carries the AM away from the NS, and spins the NS down.
One problem related to the propeller model for the torque reversals
in 4U1626$-$67 might be that X-ray pulsations were detected during
the spin-down stage, indicating that accretion was still going on
even when $r_0>r_{\rm c}$, though in simulations by \citet{sut06}
and \citet{rom04,rom05} both accretion and spin-down were observed
at the propeller stage. \citet{per06} suggested a model for
simultaneous accretion and ejection around magnetized NSs. When the
spin axis of an  NS is not aligned with the magnetic axis, the inner
radius of the disk relies on the tilt angle $\theta$ between the two
axes and the longitude $\phi$, so that the inner edge of the disk is
not circular. For certain values of $\theta$, the inner disk radius
is partially larger than $r_{\rm c}$ where the propeller mechanism
starts up. Meanwhile, the NS accretes in the region where the inner
disk radius is smaller than $r_{\rm c}$. In other words, the system
can undergo the propeller and accretion phases at the same time.
Moreover, a fraction of the ejected material does not receive enough
energy to be completely unbind, and hence falls back into the disk.
Thus it is possible that for a given accretion rate of the NS, there
are multiple solutions of the mass flow rate through the disk. The
spin evolution is determined by the AM transferred from the disk to
the NS through accretion, and that given by the NS to the ejected
matter. When $\theta$ is larger than a critical value, the system
may settle in a limit cycle of spin-up/spin-down transitions for a
constant value of the mass accretion rate.

\subsection{Bi-state model}

The general idea of the bi-state model is that there may exist two
stable states of the accretion disk corresponding to the spin-up and
spin-down of the NS. If the system is triggered to jump from
one state to the other, it appears as the torque reversal.
How to establish the two states is the essential problem for this kind of model.
In the model suggested by \citet{yi97}, the torque reversals are
caused by alternation between a Keplerian, thin disk and a
sub-Keplerian, advection-dominated accretion flow (ADAF) with small
changes in the accretion rate. When $\dot{M}$ becomes smaller than a
critical value $\dot{M}_{\rm cr}$, the inner part of the accretion
disk may make a transition from a Keplerian, thin disk to a
sub-Keplerian ADAF, in which the angular velocity in the disk
$\Omega (r)=A \Omega_{\rm K} (r)$ with $A < 1$ \citep{nar95}. In
this case the corotation radius becomes $r'_{\rm c} = A^{2/3}r_{\rm
c}<r_{\rm c}$ (Here we use the prime to denote quantities in ADAF).
The torques exerted on the NS by a Keplerian and sub-Keplerian disk
can be estimated to be
\begin{equation}
N=\frac{7}{6}N_0\frac{1-(8/7)(r_0/r_{\rm c})^{3/2}}{1-(r_0/r_{\rm
c})^{3/2}},
\end{equation}
and
\begin{equation}
N'=\frac{7}{6}N'_0\frac{1-(8/7)(r'_0/r'_{\rm
c})^{3/2}}{1-(r'_0/r'_{\rm c})^{3/2}},
\end{equation}
respectively, where
$N'_0=\dot{M}(GMr'_0)^{1/2}$. Since the inner radius of the disk
does not change much ($r_0\simeq r'_0$), the dynamical changes in
the disk structure may lead to the slow ($\omega_{\rm s}<\omega_{\rm
c}$, spin-up) and rapid ($\omega'_{\rm s}>\omega_{\rm c}$,
spin-down) rotator stage alternatively when $\dot{M}$ varies around
$\dot{M}_{\rm cr}$.

\citet{love99} developed a model for magnetic, propeller-driven
outflows that cause a rapidly rotating magnetized NS accreting from
a disk to spin-down.  An important feature of their results is that
the effective Alf\'ven radius $r_{\rm A}$ depends not only on
$\dot{M}$ and $B$, but also on $\Omega_{\rm s}$. Because $r_{\rm A}$
decreases as $\Omega_{\rm s}$ decreases, there exists a minimum
value of $\Omega_{\rm s}$ for stable accretion disks, and for a
given $\Omega_{\rm s}$, there could be two values of $r_{\rm A}$,
one larger than $r_{\rm c}$ and the  other smaller than $r_{\rm c}$.
This points to a mechanism for the propeller from being ``on" to
being ``off", when there is a change between the two possible
equilibrium configurations, leading to transitions between spin-down
and spin-up with roughly similar rates for nearly constant
$\dot{M}$. Since the transitions may be stochastic, and triggered by
small variations in the accretion flow or in the magnetic field
configuration, this model, similar as \citet{tor98}, could be
responsible for the torque reversals in Cen X-3 rather 4U 1626$-$67,
provided that  $r_{\rm A}\sim r_{\rm co}$.

\citet{loc04} presented a disk-magnetosphere interaction model where
the extent of the magnetosphere is determined by balancing the
outward diffusion and inward advection of the stellar magnetic field
at the inner edge of the disk. They showed that the
disk-magnetosphere system has two stable torque states for certain
combinations of the magnetic Prandtl numbers and the fastness
parameter. If the star is initially spinning up, in the absence of
extraneous perturbations, the spin-up equilibrium eventually
vanishes and the star subsequently spins down. In its current form,
the model does not exhibit repeated torque reversals observed.

\section{QPOs in 4U 1626$-$67 and constraints on previous models}

The mHz QPOs in 4U 1626$-$67 have been detected with {\em Ginga}
\citep{shi90}, {\em ASCA} \citep{ang95}, {\em BeppoSAX}
\citep{owe97}, {\em RXTE} \citep{kom98,cha98} and {\em XMM-Newton}
\citep{kra07}. More recently, \citet{kau08} investigated the
evolution of the QPO frequency in 4U 1626$-$67 over a long period.
It was shown that the QPO frequency in 4U 1626$-$67 during the last
22 years evolved from a positive to a negative trend: in the earlier
spin-up era, the QPO central frequency increased from $\sim 36$ mHz
in1983 to $\sim 49$ mHz in 1993, while in the subsequent spin-down
era, it gradually decreased at a rate $\sim (0.2\pm 0.05)$
mHz\,yr$^{-1}$. However, the lack of observations around 1990 does
not allow to define an exact time when the evolutionary trend of the
QPO frequency changed. It seems to be somewhat coincident with the
torque reversal.

In accretion-powered X-ray pulsars, the QPO frequency is usually
regarded to be directly related to the inner radius $r_0$ of the
accretion disk. The widely adopted  QPO theories are the Keplerian
frequency  model (KFM) and the BFM, which consider the QPO frequency
as the Keplerian frequency $\nu_{\rm K}$ at $r_0$ and the beat
between $\nu_{\rm K}$  at $r_0$ and the spin frequency $\nu_{\rm s}
(=\Omega_{\rm s}/2\pi)$
of the NS, respectively. In the case of 4U 1626$-$67, we adopt BFM
rather KFM as the proper interpretation, since the spin frequency
is larger than the QPO frequency, which means the system would be in
the propeller phase if we consider the QPO frequency to be the Keplerian
frequency at $r_0$.

According to BFM, the frequency at $r_0$ is
\begin{equation}
\nu_{\rm K}=\nu_{\rm QPO}+\nu_{\rm s},
\end{equation}
and the fastness parameter is
\begin{equation}
\omega_{\rm s}\equiv\frac{\nu_{\rm s}}{\nu_{\rm K}}
=\frac{\nu_{\rm s}}{\nu_{\rm QPO}+\nu_{\rm s}}.
\end{equation}
If the BFM correctly explains the QPOs in 4U 1626$-$67, we can infer
that $\nu_{\rm K}$ increased during the spin-up era and then
slightly decreased during the spin-down era. Since the the X-ray
luminosity $L_{\rm X}$ originating from disk accretion positively
correlates with $\nu_{\rm K}$, one would expect a similar
evolutionary trend in $L_{\rm X}$.  This is opposite the fact that
X-ray luminosity of 4U 1626$-$67 has been decreasing since 1977
\citep{kra07}. One intermediate implication is that the observed
change in the X-ray luminosity is not due to a change of mass
accretion rate by the same factor \citep{kau08}. For example, the
accretion flow in 4U1626$-$67 might compose multi-components (e.g.
disk and coronal streams), and disk accretion accounts for only a
fraction of the X-ray luminosity, but influences the evolution of
the QPO frequency. Alternatively, the X-ray luminosity variation
could be due to a change in obscuration by an aperiodically
precessing warped accretion disk, as suggested for Cen X$-$3
\citep{rai08}. A line of evidence in favor of the latter suggestion
is that, since the discovery of this X-ray source over two decades
ago, the optical flux from the accretion disk has remained
essentially constant though its overall X-ray flux has declined
\citep{cha98}.

Now we use the above results to constrain current  accretion torque
models proposed for the torque reversals. The magnetically threaded
disk model always expects that $\omega_{\rm s}$ gradually increases
when the NS evolves from  spin-up to spin-down, opposite to what we
learnt from the QPO observations. In the modified-propeller model
\citep{per06} and the bi-state model of \citet{loc04}, the inner
radius of the disk is expected to change between a relatively small
and large value, corresponding to spin-up and spin-down,
respectively. If the QPOs are related to the behavior in the inner
edge of the accretion disk, a transition of the QPO frequency is
expected. In the model of \citet{yi97}, the inner radius of the disk
might not change, but the dynamical change in the disk causes the
angular velocity at the inner edge to change from $\nu_{\rm K}$ to
$A\nu_{\rm K}$. Similarly, the retrograde disk model suggest that
the QPO frequencies should vary between $\nu_{\rm K}-\nu_{\rm s}$
and $\nu_{\rm K}+\nu_{\rm s}$. All these models predict a jump of
$\nu_{\rm QPO}$ before and after the torque reversal, which seems to
be in contradiction with observations, if BFM really works for the
QPOs in 4U1626$-$67.

\section{An alternative explanation}

The above arguments suggest that the torque exerted by the accretion
disk itself may not play the sole role in accounting for the torque
reversals, and an extra spin-down mechanism seems to be required. The
latter needs to have the following properties: (1) it occurs without
requiring large decrease in the mass accretion rate (i.e., not the
propeller-driven AM loss for rapid rotators); (2) it may last for
long episodes (up to several years); (3) its spin-down torque
increases with the mass accretion rate. A potential candidate is the
stellar and disk winds (or outflows) which have not been considered
seriously in previous works.

A wind-driven spin-down mechanism for disk-accreting magnetic stars
has been widely adopted for the classical T Tauri stars (CTTS). A
large fraction of CTTS are observed to rotate at approximately
$10\%$ of the break-up speed, although they have been actively
accreting material from their surrounding Keplerian disks for $\sim
10^6-10^7$ yr \citep{ber89}.  \citet{kon91} applied the GL model to
CTTSs to explain their slow rotation. However, it was pointed out
that, when the differentially twisting angle between the star and
the disk monotonically increases, the torque exerted by the field
lines first reaches a maximum value, then decreases. This occurs
because the azimuthal twisting of the dipole field lines generates
an azimuthal component to the field, and the magnetic pressure
associated with this component acts to inflate the field, causally
disconnecting the star and the disk \citep[e.g.,][]{aly85,uzd02}.
Thus the size of the disk region that is magnetically connected to
the star is smaller, and the magnetic spin-down torque on the star
is significantly less than in the original GL model \citep{mat05a}.
\citet{mat05b} further explored the idea of powerful stellar winds
as a solution to the AM problem, and showed that stellar winds are
capable of carrying off the accreted AM, provided that the ratio of
the outflow rate and the accretion rate is $\sim 0.1$. In this model
a significant part of the disk matter is launched as the stellar
winds, although the mechanism is unknown. If it is due to the impact
of plasma on the stellar surface from magnetospheric accretion
streams, the work by \citet{cra08,cra09} suggested that it could
produce T Tauri-like mass-loss rates of at least 0.01 times the
accretion rate.

Outflows from the disk-magnetosphere boundary were investigated by
many authors, both theoretically and numerically, and episodes of
field inflation and outflows were observed
\citep[e.g.][]{good97,good99,sut06,rom09}.
The maximum velocities in
the outflows are usually of the order of the Keplerian velocity of
the inner region of the disk. This favors the models where the
outflows originate from the inner region of the disk or from the
disk-magnetosphere boundary. In the most recent simulations,
\citet{rom09} reported that,  in the case of slowly rotating stars,
the magnetic flux of the star can be bunched up by the disk into an
X-type configuration, leading to a conical wind\footnote{In their
simulations Romanova et al. (2009) did not take into account
possible stellar wind.}, when the turbulent magnetic Prandtl number
(the ratio of viscosity to diffusivity) $ > 1$ and when the
viscosity is sufficiently high, $\alpha \ga 0.03$. The amount of
matter flowing into the conical wind was found to be \ $\sim
10-30\%$ of the disk accretion rate.

Observationally there also exists possible evidence for the
existence of  winds from 4U1626$-$67. With {\em Chandra} observation
\citet{sch01} resolved the Ne/O emission line complex near 1 keV
into Doppler pairs of broadened ($\sim 2500$ kms$^{-1}$ FWHM) lines
from highly ionized Ne and O, and suggested that they might
originate in a disk wind driven from the pulsar's magnetopause. The
wind mass loss rate $\sim 10^{-10} M_{\sun}$yr$^{-1}$ was shown to
be of the same order as the observed mass accretion rate onto the
NS. The structure of the emission lines and the
helium-like Ne IX and O VII triplets support the hypothesis that
they are formed in the high-density environment of an accretion disk
\citep{kra07}.

In an evolutionary view, perhaps winds (or outflows) are inevitably
required for 4U1626$-$67. It is related to the puzzle that the
observationally inferred mass transfer rate of $\dot{M}\sim
~2\times10^{-10}\,M_{\sun}$yr$^{-1}$ is much larger than theoretical
expectations $\dot{M}\sim ~3\times10^{-11}\,M_{\sun}$yr$^{-1}$ for
mass transfer from a $\sim 0.02 M_{\sun}$ donor in a 42 minute
binary driven by AM loss via gravitational radiation
\citep{cha98}. The discrepancy between the measured and predicted
mass transfer rates indicates that there must be other driving
mechanisms besides gravitational radiation, and wind mass loss is
one of the most suitable choices.

We now propose a model for the torque reversals based on the
wind/outflow-assisted spin-down mechanism. When there is no (or
weak) wind from the star and the disk, the total torque exerted on the
NS comes from the accreting material and the magnetic field-disk
interaction. The latter, however, may not contribute significantly
to the torque for slow rotators, according to \citet{mat05a}, so the
NS experiences a spin-up torque,
\begin{equation}
N_{\rm su} \simeq N_0=\dot{M}(GMr_0)^{1/2}.
\end{equation}
During the spin-up stage, the increase of the mass transfer rate
(observed as the increase of the QPO frequency) can lead to the
bunching of the field lines if the inward flow is faster than
outward diffusion of the field lines \citep{rom09}. When the field
topology around the magnetosphere becomes open, strong stellar $+$
disk winds are launched, and the NS enters the spin-down stage. The
total torque becomes
\begin{equation}
N_{\rm sd}= N_0+N_{\rm dw}+N_{\rm sw},
\end{equation}
where $N_{\rm dw}$ and $N_{\rm sw}$ are the torques from the disk
wind and the stellar wind, respectively. Calculations by
\citet{rom09}
 showed that close to the inner boundary of the accretion
disk a significant part of the AM within the disk matter is carried
away by the conical wind. Here we assume $N_{\rm dw}= -\chi N_0$
with $\chi$ lying between 0 and 1.
The torque from the stellar winds is
\begin{equation}
N_{\rm sw}=-\kappa\dot{M}_{\rm w}r_{\rm A}^2\Omega_{\rm s},
\end{equation}
where $r_{\rm A}$ is the Alfv\'en radius, at which the poloidal wind
velocity equals the poloidal Alfv\'en speed.  The dimensionless
factor $\kappa$ takes into account the geometry of the wind and is
order of unity ($\kappa=2/3$ for a spherically symmetric wind). The
magnitude of $r_{\rm A}$ depends on the the magnetic field strength
and geometry, mass loss rate, and wind speeds. A semi-analytic,
fitted formulation was suggested by \citet{mat08} from
two-dimensional (axisymmetric) MHD simulations,
\begin{equation}
\frac{r_{\rm A}}{R}=K(\frac{B^2R^2}{\dot{M}_{\rm w}v_{\rm esc}})^m,
\end{equation}
where $K\simeq 2.11$, $m\simeq 0.223$, $R$ is the stellar radius,
$B$ the surface magnetic field strength, and $v_{\rm esc}=
(2GM/R)^{1/2}$ the escape speed from the stellar
surface, respectively. If the wind is launched at the boundary of
the magnetosphere, then $R$ should be replaced by the
magnetospheric radius.
Adopt typical values of the parameters of 4U1626$-$67, i.e., $M=1.4
M_{\sun}$, $R=10^6$ cm, $B=3\times 10^{12}$ G, $\dot{M}_{\rm w}
\sim 0.1\dot{M}\sim 10^{15}$ gs$^{-1}$, one can obtain $r_0\simeq
5.5\times 10\times 10^8$ cm, and $r_{\rm A}/r_0\sim 1.5-6$.
Assume $\chi\sim 0.5$, the ratio of the spin-down and spin-up torques
can be estimated to be
\begin{equation}
|\frac{N_{\rm sd}}{N_{\rm su}}|\simeq 1.35(\frac{\kappa}{0.5})
(\frac{\dot{M}_{\rm w}/\dot{M}}{0.1})(\frac{r_{\rm A}/r_0}{6})^2
(\frac{\omega_{\rm s}}{0.75})-0.5,
\end{equation}
and we get $N_{\rm su}\sim -N_{\rm sd}\sim 1\times 10^{33}$
gcm$^2$s$^{-2}$, corresponding to a spin-up/down rate of $\sim (\pm)
8\times 10^{-13}$ Hz\,s$^{-1}$, compatible with the observed values.
Note that in the current model, the mass transfer rate during the
spin-down stage can maintain a higher value compared to that during
the spin-up stage, as shown by the QPO frequency evolution.
Additionally, Eqs.~(13) and (14) indicate that the spin-down rate
can increases with mass transfer rate, if a roughly constant
fraction of the transferred mass goes into the winds.

\section{Discussion and conclusions}

Based on the measurements of QPOs in 4U1626$-$67 and the beat
frequency interpretation, we proposed a model for the torque
reversals in this source. The essential idea is that the spin-down
is induced by stellar and disk winds (or outflows) that take away
the AM of the NS. Thus a significant decrease of the mass transfer
rate and possible propeller effect are not required. Because of wind
mass loss, the accretion rate of the NS is not simply the mass
transfer rate through the accretion disk. Since the latter
determines the QPO frequencies, it is not expected that there exists
straightforward, positive correlation between the QPO frequency and
the X-ray luminosity (or the spin changing rate) \citep{boz09}.

The model seems to be in line with observations of other X-ray
pulsars besides 4U 1626$-$67. For example, 4U 1907$+$09 was found to
switch from spin-down to spin-up without considerably change in
luminosity \citep{fri06}. Furthermore, \citet{int98} pointed out
that, during the spin-dwon stage, the magnetospheric radius from the
cyclotron line measurements is $r_0\sim 2400$ km, less than the
corotation radius $r_{\rm c}\sim 12 000$ km, which rules out the
possibility of the propeller effect being the spin-down mechanism.
Continuous monitoring of Her X$-$1 showed that its pulse period
evolution resembles a saw-tooth composed of spin-up and spin-down
episodes, and there occurred extremely large spin-down torques  up
to 5 times as strong as the spin-up ones, which are very likely
related to episodic ejection of matter in Her X$-$1 \citep{klo09}.
Signatures of ouflowing gas was also found in the UV spectrum of
this source \citep{vrt01,bor01}.

The main limitation of our model is the mechanism for the occurrence
of the winds, which needs to be explored in more detail. In
\citet{rom09}, strong outflows from the inner region of the disk are
expected to result from field bunching when the mass transfer rate
is enhanced, and when the magnetic Prandtl number of the turbulence
is larger than unity. The condition for the bunching of the field
lines is however, not well understood. It requires that the speed of
the inward flow of matter in the disk should be higher than the
speed of outward diffusion of the stellar field lines. If the
accretion rate is determined not only by the viscosity but by any
other mechanisms of outward AM transport, such as by the spiral
waves, then this condition will be satisfied and the bunching of
field lines is expected. The stellar winds might be driven by some
fraction of the accretion power in the way of accretion shocks
and/or magnetic reconnection events \citep[e.g.][]{mat05b,cra08}. We
also note that there could be outflows caused by heating of hard
X-ray emission of the NS, although more likely in the case of
spherical accretion \citep{ill90}. These authors suggested that, if
the X-ray luminosity falls in the region of $\sim 2\times 10^{34}$
ergs$^{-1} <L_{\rm X}<\sim 3\times 10^{36}$ ergs$^{-1}$, Compton
scattering heats the accreted matter anisotropically, and some of
the heated matter with a low density can flow up and form outflows
to take the AM away. It is interesting to note that the luminosity
of 4U 1626$-$67 at torque reversals just fulfills the criteria if
its distance is $\sim 5$ kpc.

The occurrence and disappearance of the stellar and disk winds may
be accompanied with possible change in the structure of the
magnetosphere and the accretion disk, leading to variation in the
radiation features of the NS and the disk. Observations with {\em
Chandra} and {\em XMM-Newton} showed that the pulse profile of 4U
1626$-$67 has changed significantly from what was found prior to the
torque reversal in 1990, suggesting a change in the geometry of the
accretion column \citep{kra07}. The X-ray continuum spectrum was
also shown to be closely correlated with the torque state. During
the 1977$-$1990 spin-up phase, the spectrum was well described by an
absorbed blackbody, a power law and a high energy cutoff
\citep{pra79,kii86}. After the torque reversal in 1990, the
time-averaged X-ray spectrum was found to be relatively harder
\citep{owe97,yi97,kra07}, which is often regarded as an indication
of outflows. The spectrum was also found to be harder during the new
torque transition in 2008 than before or after. These results imply
that the torque reversal is not a simple case of change in the mass
accretion rate, but there is also a change in the accretion geometry
in the vicinity of the NS.

When the inner disk comes closer to the star, there is a higher
difference in angular velocity between inner disk and magnetosphere,
and inflation of the field lines is more efficient. The difference
between the angular velocities can lead to cyclic evolution of the
field lines - development of the toroidal field component, field
line opening, and reconnection, which were suggested to be
accompanied with energy release or flaring activities
\citep[e.g.][]{aly85}. 4U 1626$-$67 was indeed seen to flare
dramatically in both X-ray and optical on timescales of $\sim 1000$
s before 1990 \citep{jos78,mcc80,li80}. However, there were no
flaring events seen in any of the observations by \citet{kra07}. The
cessation of flaring activity may have occurred at the same time as
the torque reversal. As the NS has entered the spin-up phase since
2008, it is interesting to see whether flaring will appear again.

\begin{acknowledgements}
We are grateful to the referee, Dr. Marina Romanova for comments
and suggestions that greatly helped improve the manuscript.
This work was supported by the Natural Science Foundation of China
(under grant number 10873008) and the National Basic Research
Program of China (973 Program 2009CB824800).
\end{acknowledgements}

\end{document}